\documentclass{article}

\usepackage{listings}       % For code listings
\usepackage{times}
\usepackage{graphicx}       % For including graphics
\usepackage{amsmath, amssymb} % AMS math packages
\usepackage{booktabs}       % For professional-quality tables
\usepackage{subcaption}     % For subfigures and subcaptions

\usepackage[utf8]{inputenc} % Allow UTF-8 input
\usepackage[T1]{fontenc}    % Use 8-bit T1 fonts
\usepackage{hyperref}       % For hyperlinks
\usepackage{url}            % Simple URL typesetting
\usepackage{amsfonts}       % Blackboard math symbols
\usepackage{nicefrac}       % Compact symbols for 1/2, etc.
\usepackage{microtype}      % Microtypography
\usepackage{xcolor}         % Colors

\usepackage{iclr}
\iclrfinalcopy

\title{Relevance Isn't All You Need: 

Scaling RAG Systems With Inference-Time Compute Via Multi-Criteria Reranking}

\author{
Will LeVine$^{\dag}$\thanks{Corresponding author email: \href{mailto:levinewill@icloud.com}{levinewill@icloud.com}} \\
Microsoft
\And
Bijan Varjavand\thanks{These authors contributed equally.} \\
Scale AI
}

\date{}

\begin{document}
\maketitle

\begin{abstract}
Modern Large Language Model (LLM) systems typically rely on Retrieval Augmented Generation (RAG) which aims to gather context that is useful for response generation. These RAG systems typically optimize strictly towards retrieving context that is maximally relevant to the query. However, conventional theory suggests that retrieval systems which seek to maximize context relevance without any additional explicit criteria can create information bottlenecks. We reaffirm this finding in the modern age of LLM's by showing that in standard RAG pipelines, \textit{maximizing for context relevance alone can degrade downstream response quality}. In response, we show evaluations of existing RAG methods which account for both context relevance \textit{and} answer quality. These evaluations introduce a novel finding that existing RAG systems scale poorly with inference time compute usage when considering our combined metric. We introduce "RErank BEyond reLevance (\textbf{REBEL})", which enables RAG systems to scale with inference-time compute via injection of multi-criteria optimization using Chain-of-Thought prompting (and optionally Multi-Turn dialogue). Ultimately, this enables a new performance/speed tradeoff curve, where RAG systems are able to achieve both higher relevance of retrieved contexts \textit{and} superior answer quality as inference time increases. \footnote{Code for the implementation of our method in \texttt{llama-index} can be found at the following PR: \href{https://github.com/run-llama/llama_index/pull/17590}{https://github.com/run-llama/llama\_index/pull/17590}} \footnote{Code for running experiments using this \texttt{llama-index} implementation can be found at \href{https://github.com/microsoft/REBEL}{https://github.com/microsoft/REBEL}.}
\end{abstract}

\section{Introduction}
Large Language Models (LLMs) have significantly advanced the field of natural language processing, enabling a wide range of applications from text generation to question answering. However, these models often rely solely on knowledge embedded in datasets involved during training, which limits their ability to generate responses informed by dynamic, fine-grain, or recent information. Retrieval-Augmented Generation (RAG) has emerged as a transformative paradigm to address this limitation. In a typical RAG workflow, relevant external documents are retrieved and added to a generative model's input context. This integration enhances the utility of LLMs across diverse applications, from customer support to academic research, by grounding their outputs in up-to-date and context-specific knowledge. See Appendix Figure \ref{fig:overall_rag_pipeline} for a high-level overview of a typical RAG pipeline.

A key challenge in RAG systems lies in selecting which documents to retrieve and how to rank them effectively. While many systems prioritize maximizing relevance, our findings demonstrate that doing so without considering secondary criteria leads to a tradeoff: methods that optimize solely for relevance may boost context relevance yet degrade the overall quality of the generated answer. For instance, our experiments show that while Cohere and LLM Rerank achieve high retrieval relevance, they do so at the expense of answer quality. These observations build upon results reported in works such as \citet{eibich2024aragog}, where the highest-performing RAG systems in terms of retrieval relevance often exhibited the lowest answer quality, and various modifications to RAG pipelines uniformly increased the retrieval relevance while decreasing answer quality for all RAG pipelines evaluated. This aligns with theoretical results from information theory \citep{tishby1999information}, multi-criteria decision making \citep{figueira2005multiple}, and information retrieval \citep{robertson1977probability}, and also aligns with more classical multi-criteria information retrieval methods such as Maximum Marginal Relevance \citep{carbonell1998use}, xQuAD \citep{santos2010xquad}, and PM-2 \citep{dang2013pm2}. Our experiments reaffirm in the modern age their earlier findings that optimizing for a single criterion (like relevance) can create information bottlenecks and fail to capture important properties of optimal solutions. See Appendix~\ref{sec:theoretical_foundations} for detailed analysis of the theoretical foundations.

  \begin{figure}[t]
    \centering
    \includegraphics[width=0.49\linewidth]{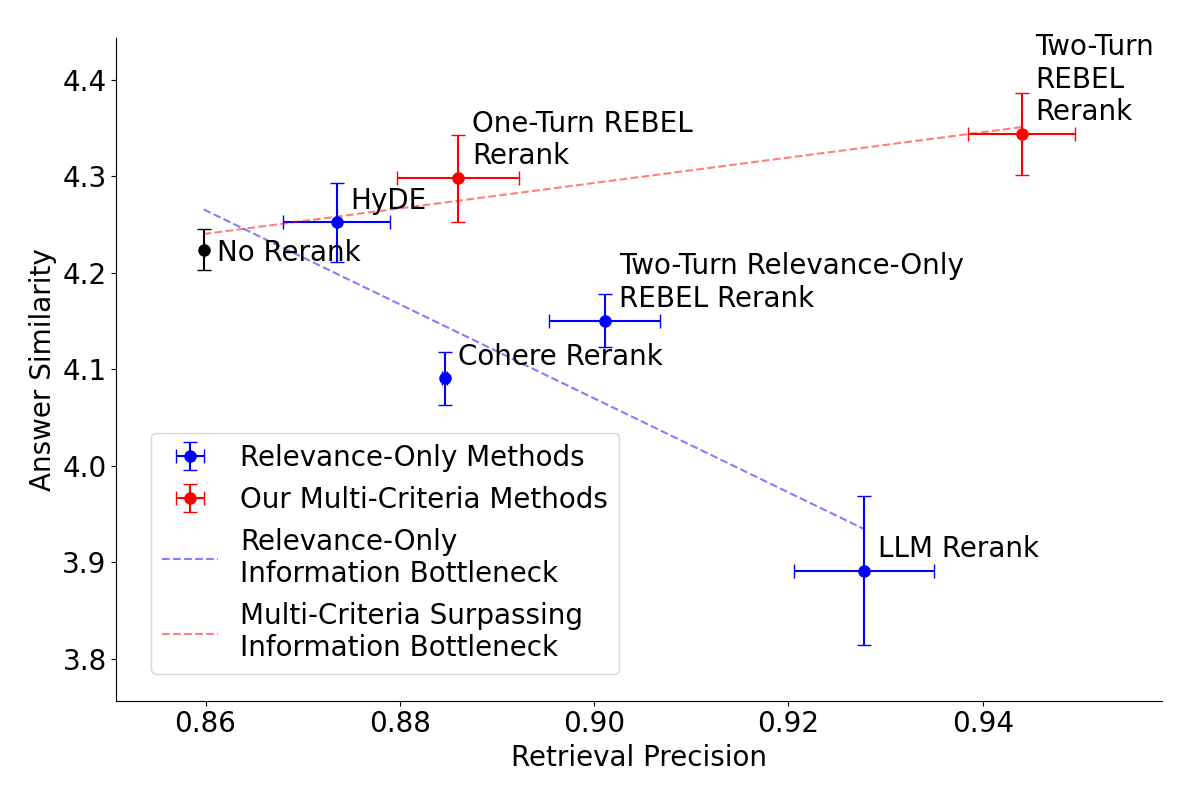}
    \includegraphics[width=0.49\linewidth]{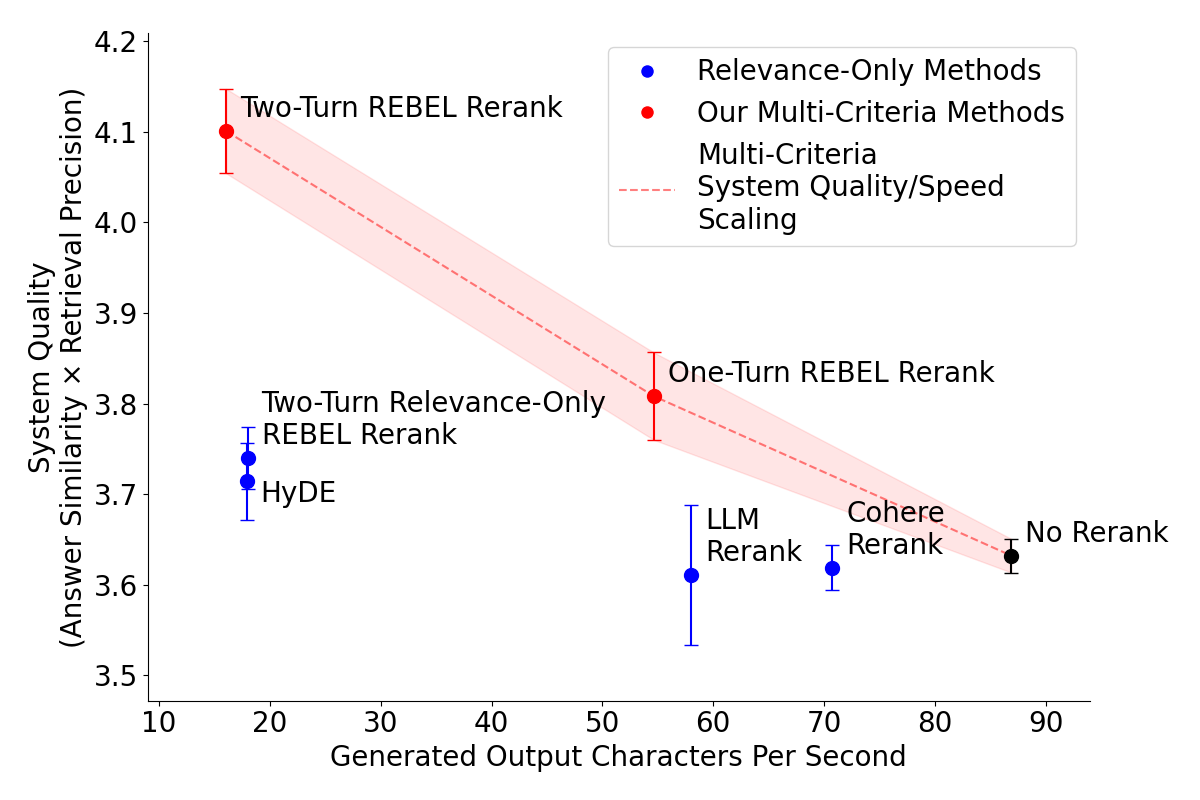}
    \caption{(\textbf{Left}) Comparison of retrieval methods showing retrieval precision versus answer similarity, with error bars indicating 95\% confidence intervals. The dashed best-fit lines represent the previously posited information bottleneck (blue) and the surpassing of that bottleneck by our multi-criteria rerankers (red). The one-turn version uses five fixed criteria (depth, diversity, clarity, authoritativeness, and recency) to achieve both higher retrieval relevance and answer quality than vanilla RAG (No Rerank). The two-turn version further improves performance by adapting criteria to each query through a two-turn prompting process. (\textbf{Right}) Visualization of system quality (measured by the multiplication of answer similarity and retrieval precision) and system inference speed (measured by generated output characters per second) for each method. We note that existing relevance-only methods are not able to achieve higher system quality at efficient inference speed rates, while our multi-criteria methods enable a new RAG tradeoff curve where inference compute can be leveraged to greatly increase system quality.}
   \label{fig:results}
\end{figure}

In contrast, our one-turn multi-criteria reranker defies the conventional relevance/quality tradeoff by incorporating secondary criteria without significant additional inference speed, as compared to LLM Rerank and Cohere Rerank. Moreover, we show that query-dependent selection of secondary criteria allows further improvements at the cost of additional inference time.

Our contributions are as follows:
\begin{enumerate}
    \item \textbf{Single-Criterion Relevance/Answer Quality Tradeoff Demonstration:} We reaffirm in the modern age of LLM's the prior mentioned theoretical foundations that posit the tradeoff between relevance and answer quality when secondary criteria are ignored. Specifically, we demonstrate that methods optimizing solely for relevance, such as Cohere and LLM Rerank, achieve higher retrieval precision while significantly degrading answer quality.
    
    \item \textbf{One-Turn Multi-Criteria Reranking:} We show that a one-turn multi-criteria LLM reranking prompt can defy this relevance/quality tradeoff. By measuring secondary qualities that are essential for evaluating context to an LLM - in addition to relevance - our one-turn approach is able to achieve both higher relevance of retrieved contexts \textit{and} higher answer similiarity as compared to a system with no reranking.
    
    \item \textbf{Two-Turn Multi-Criteria Strategy:} We introduce a two-turn meta-prompting strategy that dynamically infers query-dependent criteria, leading to the highest answer quality and context relevance at the cost of additional inference speed.

    \item \textbf{New Inference-Time-Compute/Quality Tradeoff Curve In RAG Systems:} Ultimately, along with no reranking, our one-turn and two-turn methods enable a new tradeoff curve in RAG systems, where there now exists a trade off between inference speed and overall system quality as measured by both increased context relevance \textit{and} improved answer quality.
\end{enumerate}

\begin{figure}[t]
    \centering
    \includegraphics[width=1\linewidth]{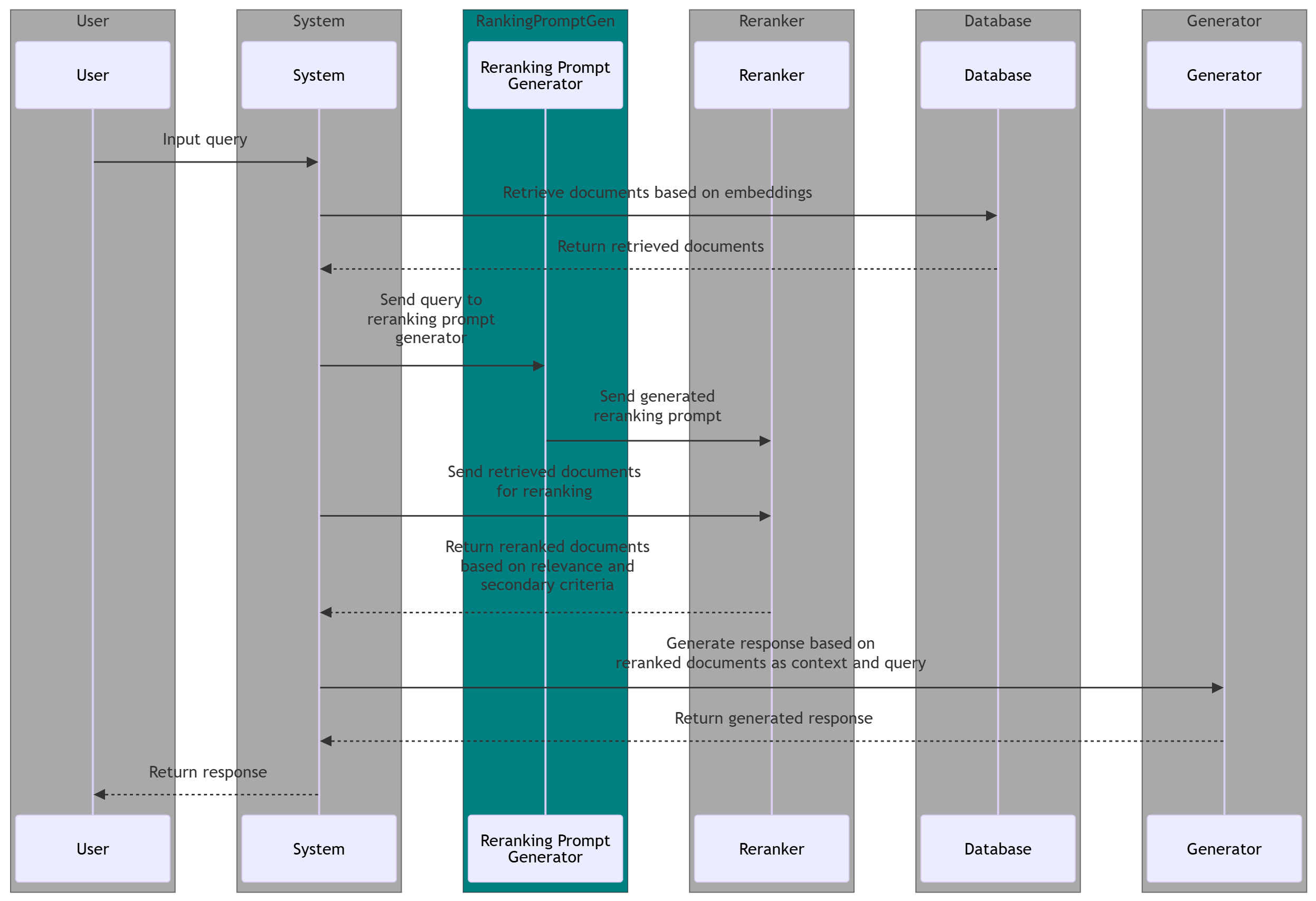}
    \caption{The two-turn version of REBEL Rerank enhances RAG systems by generating query-dependent reranking prompts that guide document selection based on both relevance and secondary criteria (such as authoritativeness, diversity, and recency) inferred from the user query. The Reranking Prompt Generator creates custom prompts that help the Reranker evaluate retrieved documents using a comprehensive scoring system that extends beyond simple relevance matching. Our experiments show that this approach maintains high retrieval relevance while significantly improving end-to-end answer quality, challenging the conventional assumption that maximizing relevance alone leads to optimal results. This finding suggests that the quality of RAG-generated responses depends not just on the topical relevance of retrieved documents, but on a broader set of contextual criteria that vary by query type and domain.}
    \label{fig:our_approach}
\end{figure}

\section{Our Method}
REBEL introduces two complementary approaches that incorporate chain-of-thought prompting into the reranking process. First, a one-turn multi-criteria reranking prompt measures qualities in addition to relevance to defy the conventional relevance/quality tradeoff. Second, a dynamic two-turn strategy adapts to individual queries by inferring query-specific criteria.

\subsection{One-Turn Multi-Criteria Strategy}
The foundation of our one-turn approach is a fixed prompt that instructs the reranking LLM to evaluate documents based on predefined secondary criteria in addition to mere topical relevance:

\begin{enumerate}
    \item \textbf{Secondary Criteria Evaluation:} The prompt defines a set of criteria beyond basic relevance that capture qualities essential for LLM context evaluation:
    \begin{itemize}
        \item \textbf{Depth of Content:} Measuring thoroughness and comprehensive coverage of the topic.
        
        \item \textbf{Diversity of Perspectives:} Evaluating the representation of multiple viewpoints or angles.
        
        \item \textbf{Clarity and Specificity:} Assessing how clearly and precisely the document presents information and addresses the query.
        
        \item \textbf{Authoritativeness:} Measuring source credibility and expertise.
        
        \item \textbf{Recency:} Evaluating temporal relevance.
    \end{itemize}

    \item \textbf{Comprehensive Scoring Rubric:} The prompt provides detailed scoring guidelines:
    \begin{itemize}
        \item Relevance scores (0-10) assess topical alignment
        \item Secondary criteria scores (0-5) evaluate each additional property
    \end{itemize}

    \item \textbf{Weighted Composite Score:} Documents receive a final score computed as:
    \[
    \text{Final Score} = \text{Relevance} + \sum_i w_i \times (\text{Property}_i),
    \]
    where $w_i$ represents the weight for each secondary criterion. This weighting allows for control over how much significance is given to each component in the equation. $w_i = 0.5$ in all of our experiments. We leave tuning of these weights to future work. 

    \item \textbf{Chain-of-Thought Process:} The prompt guides the LLM through explicit reasoning steps:
    \begin{enumerate}
        \item Analyzing document content thoroughly
        \item Assigning scores independently for each secondary property
        \item Computing the weighted composite score
        \item Sorting and filtering documents based on weighted composite score
    \end{enumerate}

    \item \textbf{Strict Output Format:} The prompt enforces a consistent structure for reranking outputs that matches traditional LLM reranker formats.
\end{enumerate}

We include in Appendix Section \ref{sec:static_reranking_prompt} our one-turn multi-criteria reranking prompt for reference.

\subsection{Two-Turn Multi-Criteria Strategy}
Building on the one-turn approach, our dynamic strategy adapts the reranking criteria to each query through a two-turn process:

\begin{enumerate}
    \item \textbf{Query-Dependent Reranking Prompt Generation:} 
    The system first infers secondary criteria relevant to the user query via chain-of-thought and develops a reranking prompt. This includes:
    \begin{enumerate}
        \item Analyzing query intent and requirements
        \item Identifying which secondary criteria are most appropriate and stating their definitions in detail
        \item Specifying appropriate weighting schemes for each secondary criteria
        \item Defining the weighted composite score
        \item Instructing the reranking LLM to sort and filter documents based on the weighted composite score, with outputs adherent to the traditional LLM reranking output format.
    \end{enumerate}

    \item \textbf{Reranking:}
    The reranking LLM takes as input the generated query-dependent reranking prompt, along with a set of documnents, and produces an ordering for these documents.
\end{enumerate}

A key component of our two-turn approach is the inclusion of k-shot examples directly within the meta prompt. Specifically, we provide several sample user queries, each followed by an illustrative reranking prompt that demonstrates:
\begin{itemize}
    \item How to identify relevant secondary criteria for different types of queries
    \item How to define scoring rubrics for both relevance and secondary criteria
    \item How to formulate weighted composite scores that balance relevance with secondary criteria
    \item How to maintain consistent output formats matching typical LLM rerankers
\end{itemize}

For the complete multi-criteria meta-prompt including k-shot examples, see Appendix Section \ref{sec:meta_prompt}.

For a diagram illustrating how our two-turn method fits into a RAG system, see Appendix Figure \ref{fig:our_approach}.

\section{Evaluation Metrics and Motivation}
The evaluation of retrieval systems typically relies on metrics such as Mean Reciprocal Rank (MRR), precision@k, and recall@k \citep{manning2008introduction}, which assess relevance. While these metrics are suitable for traditional retrieval tasks, they fall short in evaluating end-to-end performance in RAG systems \citep{chen2023survey}, where the ultimate goal is to enable high-quality LLM-generated answers. The quality of these answers, our findings show, are often paradoxically degraded when RAG systems effectively retrieve highly relevant chunks if secondary criteria are ignored. This necessitates a reevaluation of how retrieval systems are measured and optimized.

\subsection{End-to-End System Evaluations}
The primary goal of our work is to improve the quality of answers generated by RAG systems. To this end, we adopt \textbf{answer similarity} \citep{tonicai2023metrics} as our primary evaluation metric. Answer similarity estimates how well the generated answer aligns with a reference answer via a rubric-based LLM - on a scale from 0 to 5. As a rubric-based metric, it provides a reliable assessment of end-to-end system performance - as has been shown in \citet{kim2023prometheus} which reports that LLM evaluators achieve Pearson correlations upwards of 90\% with human evaluators on rubric-based tasks.

By focusing on answer similarity, we move beyond the traditional view that RAG components to an LLM system can be evaluated in a vacuum. Instead, we assess their holistic contribution to the overall quality of generated answers. This shift aligns with the ultimate objective of RAG systems: helping LLM's deliver answers that are not only accurate but also contextually nuanced and aligned with user expectations.

For our rationale on the choice of answer similarity over alternative approaches to answer quality such as ROUGE \citep{lin2004rouge}, BLEU \citep{papineni2002bleu}, BERTScore \citep{zhang2020bertscore}, and complex multi-metric frameworks \citep{karpukhin2020dense, khattab2020colbert, lewis2020retrieve}, see Appendix Section \ref{sec:alternative_answer_quality_evaluation_metrics}.

\subsection{Balancing Secondary Criteria With Relevance}
To measure context relevance, we include \textbf{retrieval precision} \citep{tonicai2023metrics} as a secondary metric. Retrieval precision measures the proportion of retrieved contexts that are topically relevant to the query (as estimated by an LLM evaluator). Specifically, the LLM evaluator estimates for each piece of context a binary relevance score. The retrieval precision is then the average of the contexts' binary relevance scores. See Appendix Figure \ref{fig:retrieval_precision} for a more detailed view of retrieval precision. 

\subsection{Ensuring Apples-to-Apples Comparisons}
To isolate the effects of our method and avoid confounding factors, we ensure an \textbf{apples-to-apples comparison} by using the same dataset and LLM evaluator for both answer similarity and retrieval precision. This eliminates potential biases introduced by differences in evaluation models or dataset distributions. For instance, if we had instead evaluated relevance on a different dataset, this could introduce doubt that perhaps our method merely performs well on our selected answer similarity dataset across \textit{any} given metric, and the explanation for high answer similarity from our method on our evaluation dataset despite moderate relevance on a different dataset is merely due to a difference in these evaluation datasets in terms of method preference; to remove this doubt, we use the same evaluation dataset for both answer similarity and retrieval precision. We additionally use the same LLM evaluator for evaluating both generation quality (answer similarity) and context relevance (retrieval precision) - as opposed to using an LLM evaluator for one and human labels (or a different LLM evaluator) for the other. This is in order to remove doubt that perhaps our chosen LLM evaluator simply prefers answers from generations produced on contexts selected by our method across \textit{any} given evaluation metric estimated by that LLM, and our results are merely due to that LLM evaluator preference towards our method; we therefore use the same LLM evaluator (rather than human labels or a different LLM evaluator) to measure both relevance and answer similarity. 

\section{Experimental Setup}
The following setup (and textual description) is largely taken from \citet{eibich2024aragog}:

\subsection{RAG Data Sources}
This study utilizes a tailored dataset derived from the AI ArXiv collection, accessible via Hugging Face \citep{calam2023arxiv}. The dataset consists of 423 selected research papers centered around the themes of AI and LLMs, sourced from arXiv. This selection offers a comprehensive foundation for constructing a database to test the RAG techniques and creating a set of evaluation data to assess their effectiveness.

\subsubsection{RAG Database Construction}
For the study, a subset of 13 key research papers was selected for their potential to generate specific, technical questions suitable for evaluating Retrieval-Augmented Generation (RAG) systems. Among the selected papers were significant contributions such as RoBERTa: A Robustly Optimized BERT Pretraining Approach \citep{liu2019roberta} and BERT: Pre-training of Deep Bidirectional Transformers for Language Understanding \citep{devlin2018bert}. To better simulate a real-world vector database environment, where noise and irrelevant documents are present, the database was expanded to include the full dataset of 423 papers available. The additional 410 papers act as noise, enhancing the complexity and diversity of the retrieval challenges faced by the RAG system.

\subsubsection{Chunking Approach}
A \href{https://docs.llamaindex.ai/en/stable/api_reference/node_parsers/token_text_splitter/}{TokenTextSplitter} was employed with a chunk size of 2000 tokens and an overlap of 200 tokens. This approach split the documents into smaller chunks while maintaining context by allowing for overlapping text between chunks. We note that we deliberately do not use \href{https://docs.llamaindex.ai/en/stable/api_reference/packs/sentence_window_retriever/}{Sentence Window Retrieval} because, much like in the analysis of \citet{eibich2024aragog}, it uniformly decreased answer similarity in our experiments. It was then excluded for brevity. Experimental results on the effects of Sentence Window Retrieval can be founded in \citet{eibich2024aragog}.

\subsubsection{Evaluation Data Preparation}
The evaluation dataset comprises 107 question-answer (QA) pairs generated with the assistance of \texttt{GPT-4}. The generation process was guided by specific criteria to ensure that the questions were challenging, technically precise, and reflective of potential user inquiries sent to a RAG system. Each QA pair was then reviewed by humans to validate its relevance and accuracy, ensuring that the evaluation data accurately measures the RAG techniques’ performance in real-world applications. The QA dataset is available in the 
 \href{https://github.com/predlico/ARAGOG/blob/main/eval_questions/benchmark.json}{ARAGOG \citep{eibich2024aragog} Github repository} that originally proposed this experimental setup.

 For information on why we chose this dataset over more established ones, please see Appendix Section \ref{sec:why_our_dataset}.

\subsubsection{Embedding Model} In all experiments, we use OpenAI's embedding model \href{https://openai.com/index/new-embedding-models-and-api-updates/}{text-embedding-3-large} for populating our vector database.

\subsection{Mitigating LLM Output Variability}
To address the inherent variability of LLM outputs, the methodology included conducting 10 runs for each RAG technique. This strategy was chosen to balance the need for statistical reliability against the limitations of computational resources and time. Associated boxplots (including error bars) are included for full transparency into the effects of LLM output variability on the various metrics in the different RAG pipelines.

\subsection{LLM}
We use \texttt{GPT-4o} as our LLM in all inference capacities. This includes using \texttt{GPT-4o} as the LLM that takes in the meta prompt and produces reranking prompts, the LLM that takes in the reranking prompt along with the retrieved contexts and reranks the contexts, and the LLM that takes in the contexts along with the user query and produces the ultimate answer. We \texttt{GPT-4} as the LLM evaluator when calculating answer similarity and retrieval precision. We chose \texttt{GPT-4} and \texttt{GPT-4o} because of their cost-effectiveness and ease of implementation. We acknowledge that using \texttt{o1} or \texttt{o1-pro} could have led to better reranking prompts, more accurate reranking, higher quality generated answers, and more precise grading - although at a significantly higher cost. We note that we use \texttt{GPT-4} as the LLM evaluator (and deliberately do not use \texttt{GPT-4o} in this capacity) as to avoid one LLM grading its own outputs.

\section{Results}
For our experimental results, see Figure \ref{fig:results}. Information about other methods involved in these experiments can be found in Appendix Section \ref{sec:rag_techniques}. 

\subsection{Impact of Multi-Criteria Reranking}
Our experiments show that both versions of REBEL, along with no reranking, establish a new relationship that defies the information bottleneck of existing single-criteria relevance-only RAG methods. In this new relationship, answer quality increases as context relevance increases. We further note that in this new relationship, increased inference compute allows for both retrieval precision \textit{and} answer similarity to improve.

\subsection{Reaffirming The Importance of Multi-Criteria Information Retrieval In The Modern Age of LLM's}
Aside from underscoring the efficacy of our method, we also note that the plotted Relevance-Only Information Bottleneck (blue line) reaffirms past theories and findings \citep{eibich2024aragog, tishby1999information, figueira2005multiple, robertson1977probability} - now updated for the modern age of LLM's - that maximizing relevance alone can be detrimental to answer quality. REBEL's multi-criteria rerankers address this by balancing relevance with other important factors, thereby enhancing the overall utility of the generated answers.

\section{Limitations}
Since our experimental setup is largely copied from \citet{eibich2024aragog}, some of the below limitations are similar to theirs:
\begin{itemize}
    \item \textbf{Model selection:} We used \texttt{GPT-4} for evaluating responses due to the constraints of Tonic Validate, which requires the use of OpenAI models. The choice of \texttt{GPT-4}, while cost-effective, may not offer the same depth of analysis as more advanced models like \texttt{o1}.
    \item \textbf{Data and question scope:} The study was conducted using a singular dataset and a set of 107 questions, which may affect the generalizability of the findings across different LLM applications. Expanding the variety of datasets and questions could potentially yield more comprehensive insights.
    \item \textbf{Evaluation metrics:} The lack of a clear consensus on the optimal metrics for evaluating RAG systems means our chosen metrics—Retrieval Precision and Answer Similarity—are not agreed upon as the best ways to evaluate end-to-end LLM generating systems. This highlights an area for future research to solidify such evaluation framework.
\end{itemize}

\section{Future Work}
\subsection{Safety Implications and Applications}
The ability to infer and optimize for secondary criteria beyond relevance opens promising avenues for enhancing LLM safety in RAG systems. Current safety approaches often focus on model-level interventions like constitutional AI \citep{bai2022constitutional} or RLHF \citep{ouyang2022training}, but our work suggests that context selection itself can serve as an additional safety mechanism.

Specifically, REBEL could be extended to incorporate safety-focused secondary criteria such as:

\begin{itemize}
    \item \textbf{Factual Verifiability:} Prioritizing documents with clear citations, empirical evidence, or verifiable claims to reduce hallucination and misinformation risks.
    
    \item \textbf{Bias Detection:} Including criteria that assess documents for potential demographic or ideological biases, helping ensure balanced context selection.
    
    \item \textbf{Content Safety:} Evaluating documents for harmful content, extremist viewpoints, or unsafe instructions that could influence model outputs.
    
    \item \textbf{Source Credibility:} Weighting authoritative and peer-reviewed sources more heavily for sensitive topics like medical or legal advice.
\end{itemize}

This approach is particularly promising because it operates orthogonally to existing safety measures - by curating safer context, we can enhance safety regardless of the base model's training or architecture.

Furthermore, the transparency of our reranking prompts provides an auditable trail for safety-critical applications. Unlike black-box safety filters, stakeholders can inspect and modify the safety criteria being used via the query-dependent reranking prompt, enabling domain-specific safety customization. This aligns with recent calls for interpretable and controllable safety measures in AI systems \citep{weidinger2022taxonomy}. One could also imagine a version of this process which outputs not only scores, but also  justifications as to what elements of (un)desirability/(un)safeness led to the scores associated with the corresponding contexts for further transparency and explainability into how the system moderates contexts.

\subsection{Enhancing Criteria Inference through Advanced Chain-of-Thought And Multi-Turn Techniques}
Given that both versions of REBEL use Chain-of-Thought prompting, and our two-turn version uses multi-turn techniques, several promising avenues for improvement emerge from recent advances in Chain-of-Thought and multi-turn prompting. These are outlined for in Appendix Section \ref{sec:mt_improvements}.

\section{Conclusion}
We have presented \textbf{RErank BEyond reLevance (\textbf{REBEL})}, a framework that enhances retrieval-augmented generation through two complementary approaches to multi-criteria reranking. We show that incorporation of both fixed and dynamic secondary criteria beyond relevance improves RAG systems, both measured in a vacuum and as part of a larger end-to-end system. Both approaches demonstrate that optimizing for relevance alone in a RAG component of a larger end-to-end LLM system is insufficient for optimal answer generation---a finding that aligns with and empirically validates theoretical predictions about the limitations of single-criterion optimization. These innovations collectively challenge the traditional assumption that relevance alone suffices for a performant modern RAG system, paving the way for more sophisticated and effective retrieval methods - and ultimately establishing a new paradigm where inference-time compute can help RAG components of LLM systems achieve higher context relevance \textit{and} answer quality simultaneously. 

We hope that REBEL will inspire further advancements in the field.

\newpage

\bibliography{citations.bib}
\bibliographystyle{citations}

\newpage
\appendix
\onecolumn
\renewcommand{\thesection}{\Alph{section}}
\setcounter{section}{0}
\begin{center}
    \textbf{Appendix}
\end{center}

\section{Why Our Dataset?}
\label{sec:why_our_dataset}

While numerous established datasets exist for evaluating RAG systems—including Natural Questions \citep{kwiatkowski2019natural}, TriviaQA \citep{joshi2017triviaqa}, SQuAD \citep{rajpurkar2016squad}, MS MARCO \citep{nguyen2016ms}, HotpotQA \citep{yang2018hotpotqa}, and WebQuestions \citep{berant2013semantic} — we deliberately chose to work with this specialized AI ArXiv dataset and corresponding evaluation set. This choice was motivated by a critical limitation in conventional RAG evaluation datasets: their focus on factual correctness at the expense of other important qualities that influence user satisfaction with generated responses.

Traditional QA datasets typically feature concise, factual answers that primarily test a system's ability to retrieve and state correct information. For instance, Natural Questions has a median answer length of just 4 words for short answers and 40 words for long answers \citep{kwiatkowski2019natural}, and SQuAD's answers average 3.2 tokens \citep{rajpurkar2016squad}. While MS MARCO includes longer passages (most answers contain 15-40 words), its evaluation still focuses primarily on factual correctness rather than qualitative aspects of the responses.While such datasets excel at evaluating factual accuracy, they are less effective at distinguishing between systems that produce equally correct but qualitatively different responses. Specifically, when multiple RAG systems generate factually accurate answers but vary in their alignment with user preferences (e.g., in terms of explanation depth, perspective balance, or reasoning clarity), comparison against short reference answers fails to meaningfully capture these qualitative differences. In contrast, our dataset features substantially longer ground truth answers with a median length of 30 tokens, ranging from 8 to 61 tokens, with the majority of answers containing between 25-35 tokens. This increased length allows for more nuanced evaluation of response quality beyond mere factual accuracy, enabling better discrimination between systems that produce technically correct but qualitatively different responses. 

Our evaluation dataset addresses this limitation by incorporating longer, more comprehensive reference answers that better reflect the depth and nuance users typically expect. This design choice enables our evaluation metrics to better distinguish between systems that are merely factually correct and those that additionally align with user preferences for thorough, well-reasoned responses. This capability is particularly crucial for evaluating REBEL, as our method specifically aims to enhance response quality beyond basic factual accuracy by incorporating multiple criteria in the context selection process.

\section{Theoretical Foundations for Multi-Criteria Optimization in RAG}
\label{sec:theoretical_foundations}

The limitations of single-criterion optimization in retrieval systems can be understood through multiple theoretical lenses. The Information Bottleneck framework \citep{tishby1999information} demonstrates how optimizing for a single information measure can create representational bottlenecks that limit the system's ability to capture all relevant aspects of the data. In the context of RAG systems, this suggests that focusing solely on relevance may constrain the retrieval system's ability to capture other important document properties that contribute to answer quality.

This aligns with fundamental results from multi-criteria decision theory \citep{figueira2005multiple}, which establish that single-criterion optimization often fails to capture Pareto-optimal solutions in multi-objective spaces. When applied to document retrieval, this implies that Relevance-Only Single-Turn optimization may systematically exclude documents that offer better trade-offs between relevance and other crucial properties like authoritativeness or diversity.

In information retrieval theory specifically, the probability ranking principle \citep{robertson1977probability} and its extensions have highlighted the limitations of pure relevance-based ranking. These works show that document utility depends on multiple factors beyond topical relevance, particularly when documents are used as context for downstream tasks. This theoretical foundation supports our empirical finding that incorporating multiple criteria can improve end-to-end RAG system performance while maintaining strong relevance scores.

The optimality of multi-criteria approaches can also be understood through utility theory \citep{keeney1976decisions}. Just as portfolio theory demonstrates the benefits of diversification in finance \citep{markowitz1952portfolio}, RAG systems benefit from considering multiple document properties rather than optimizing for relevance alone. This diversification of criteria helps ensure the retrieved context better serves the downstream generation task.

\section{RAG Techniques}
\label{sec:rag_techniques}
The following text is taken directly from \citep{eibich2024aragog}. They include more methods, as the purpose of their paper is to cast a wide net of methods to evaluate. However, the purpose of our paper is to show the effects of including secondary criteria in reranking, and we therefore focus principally on comparing RAG systems with widely-deployed traditional LLM rerankers. We note that RAG systems in their evaluations that involved LLM rerankers were the highest perforing in terms of answer similarity anyways.

Multi-Query \citep{langchain2023query} did not significantly affect our results - experiments with Multi-Query were therefore omitted for brevity, though they are included in our GitHub repository. 

\subsection{HyDE}
The Hypothetical Document Embedding \citep{gao2022precise} technique enhances the document retrieval by leveraging LLMs to generate a hypothetical answer to a query. HyDE capitalizes on the ability of LLMs to produce context-rich answers, which, once embedded, serve as a powerful tool to refine and focus document retrieval efforts. See Appendix Figure \ref{fig:hyde} for an overview of HyDE RAG system workflow.

\subsection{Cross-Encoder}
Cross-encoders enhance RAG systems by jointly processing queries and documents to assess relevance, unlike bi-encoders which encode them separately \citep{humeau2020poly,MacKenzie2020RankingPW}. This architecture enables richer interaction between the query and document text, allowing for more nuanced relevance assessment \citep{nogueira2019passage}. Cross-encoders have shown strong performance in reranking tasks across various domains \citep{gao2021complement}, though at the cost of higher computational overhead since they must process each query-document pair. See Appendix \ref{fig:reranking} for an overview of the reranker RAG system workflow. While effective, cross-encoders typically require training or fine-tuning on domain-specific data to achieve optimal performance \citep{thakur2021beir}. 

One tool in this domain is Cohere rerank, which uses a cross-encoder architecture to assess the relevance of documents to the query. This approach differs from methods that process queries and documents separately, as cross-encoders analyze them jointly, which could allow for a more comprehensive understanding of their mutual relevance.

\subsection{LLM Rerank}
Following the success of cross-encoders in document reranking, LLM rerankers emerged as an alternative approach that leverages large language models' comprehensive language understanding capabilities for reranking retrieved documents \citep{liu2023llm}. Unlike cross-encoders which require training or fine-tuning, LLM rerankers can perform zero-shot reranking through prompts that guide their relevance assessment. While computationally more expensive than cross-encoders, LLM rerankers can potentially achieve superior accuracy by utilizing their broader knowledge and reasoning capabilities. This makes them particularly suitable for applications where reranking quality outweighs computational efficiency considerations. The workflow shown in Appendix Figure \ref{fig:reranking} also applied to LLM Rerankers.

\subsection{Two-Turn Relevance-Only REBEL Rerank}
To isolate the effects of multi-criteria optimization, we provide results for a variant of our two-turn strategy where we optimize solely for relevance. In this strategy, we update our meta prompt to instruct the reranking prompt generator to form a reranking prompt that strictly requests that the downstream reranking LLM measure relevance of documents to the query. This meta prompt does not include k-shot examples of reranking prompts. This still includes Chain-of-Thought prompting and dynamically adapting this reranking prompt to the query in terms of what to look for in order to deem a document relevant. Our full two-turn relevance-only REBEL meta prompt can be found in our Github repository.

\begin{figure}
    \centering
    \includegraphics[width=1\linewidth]{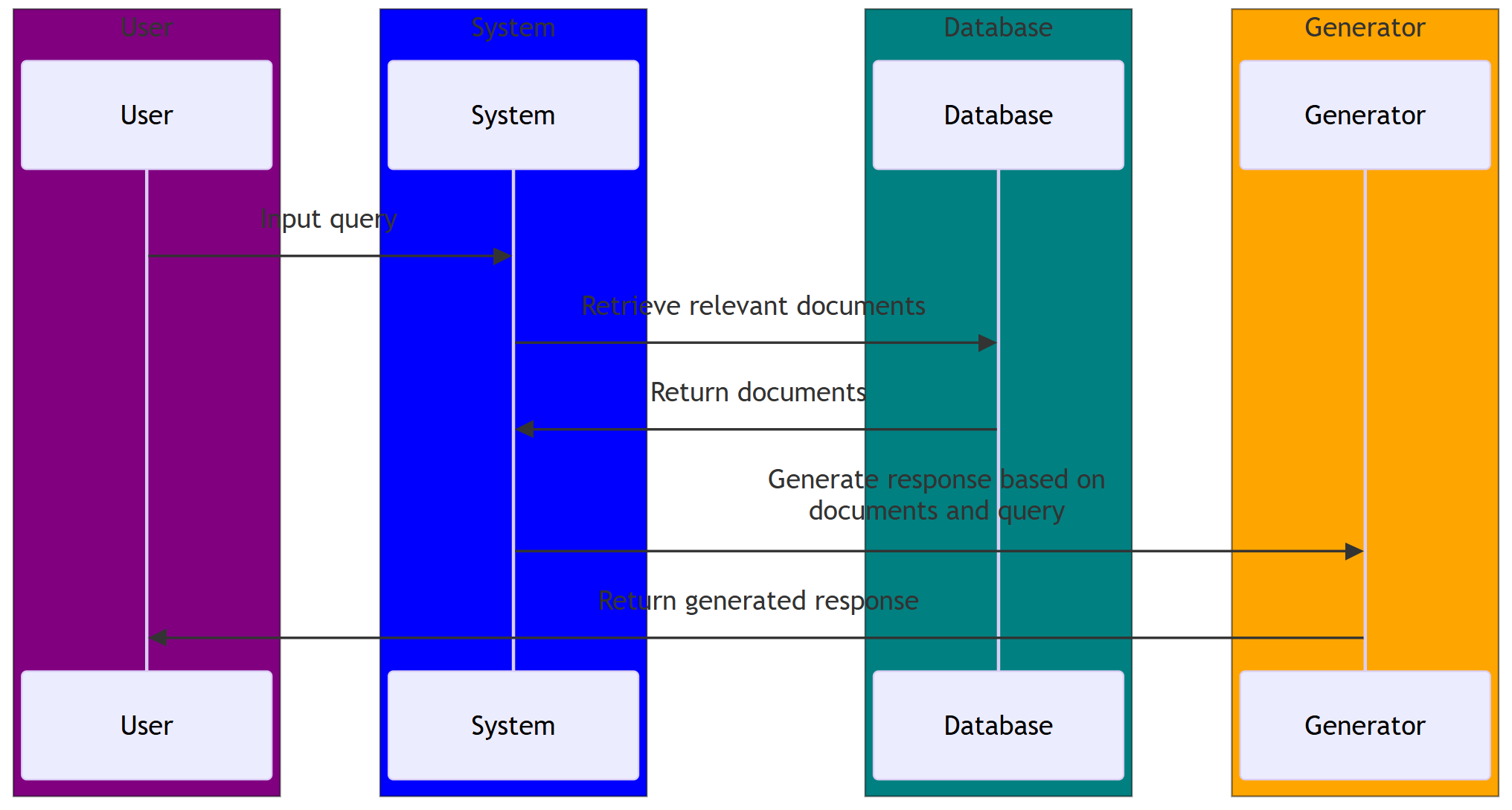}
    \caption{An overview of a Retrieval Augmented Generation (RAG) pipeline, including the usages of user queries in retrieving documents and documents in response generation. Inspired by \citet{eibich2024aragog}.}
    \label{fig:overall_rag_pipeline}
\end{figure}

\begin{figure}
    \centering
    \includegraphics[width=1\linewidth]{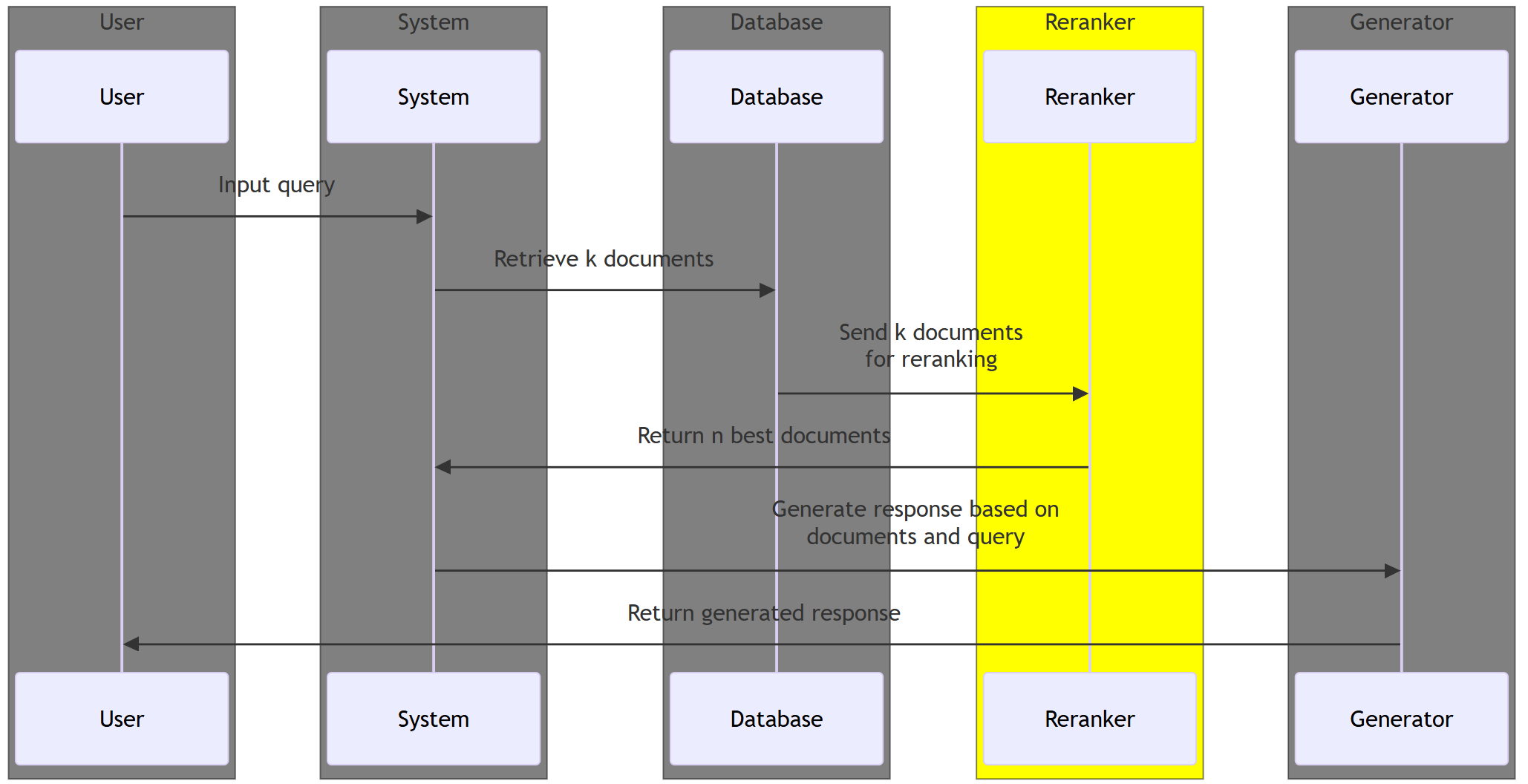}
    \caption{An overview of reranking within a RAG system. This shows how a set of $k$ retrieved documents are further refined to a set of $n$ more curated set of documents, followed by these $n$ documents being used for generation.  Inspired by \citet{eibich2024aragog}.}
    \label{fig:reranking}
\end{figure}

\begin{figure}
        \centering
        
        \includegraphics[width=\linewidth]{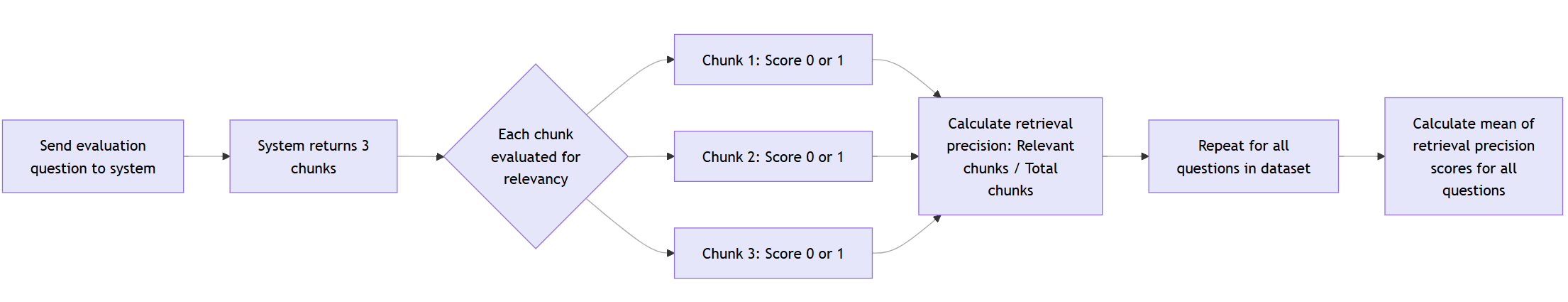}
        \caption{Detailed view of the calculation of retrieval precision. Inspired by \citet{eibich2024aragog}.}
        \label{fig:retrieval_precision}
    \end{figure}

    \begin{figure}
        \centering
        \includegraphics[width=1\linewidth]{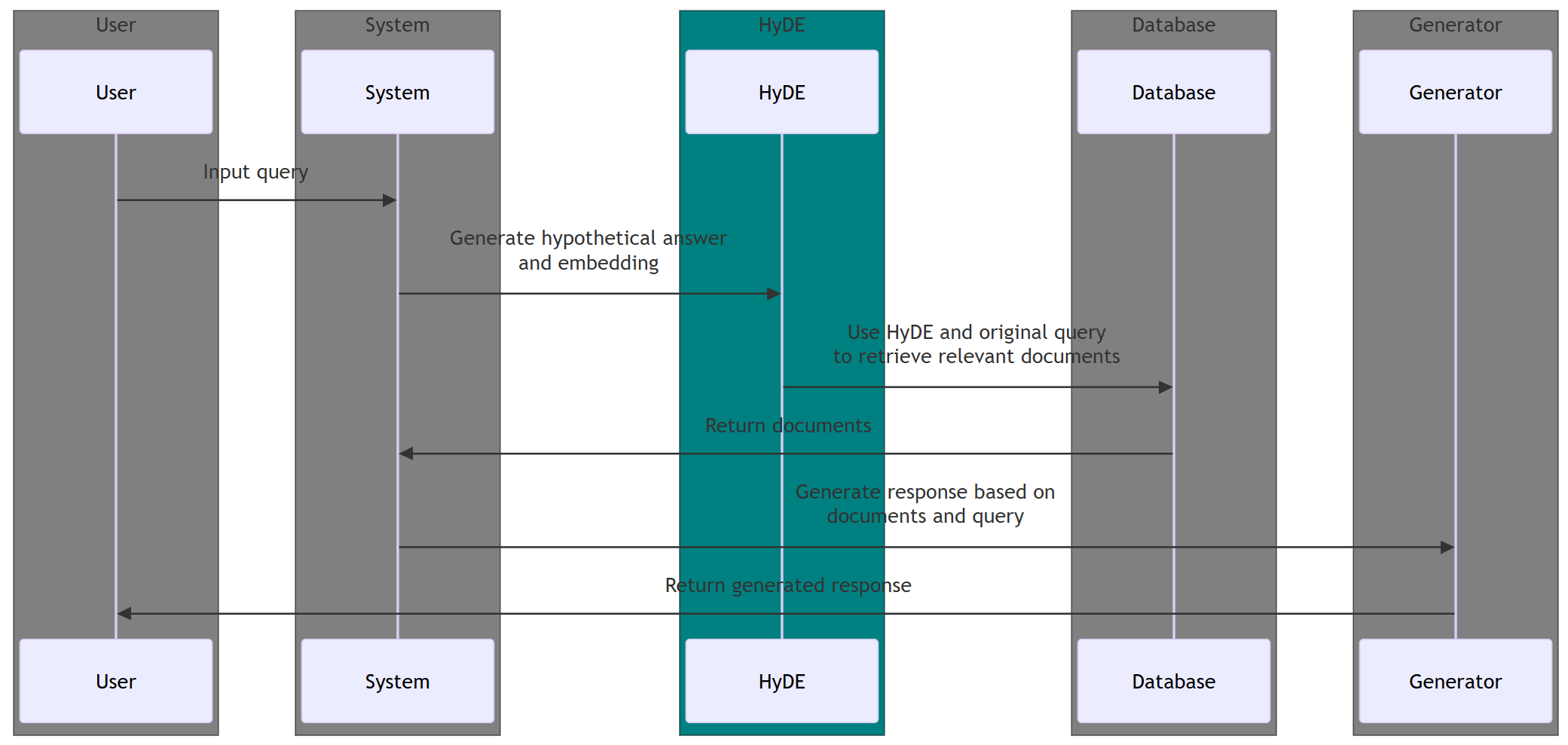}
        \caption{Process flow of the Hypothetical Document Embedding (HyDE) technique within a Retrieval Augmented Generation (RAG) system. The system takes a user query as input and leverages a Large Language Model (LLM) to generate a hypothetical answer, which is then embedded into a vector space. This embedding is used to perform vector search against a document database, retrieving relevant documents. Inspired by \citet{eibich2024aragog}.}
        \label{fig:hyde}
    \end{figure}

    \section{Multi-Criteria and Hybrid Reranking Approaches}
\label{sec:multi_criteria_methods}

This section provides an overview of existing multi-criteria and hybrid reranking approaches in information retrieval (IR). These methods have informed the development of retrieval-augmented generation (RAG) systems but have not been fully adapted to the capabilities and demands of modern large language models (LLMs).

\subsection{Learning-to-Rank (LTR) Frameworks}
Learning-to-rank (LTR) methods optimize ranking functions by using supervised learning with labeled data. Popular LTR approaches include:
\begin{itemize}
    \item \textbf{RankNet} \cite{burges2005ranknet}: A neural pairwise ranking model that uses a probabilistic cost function to learn relative document rankings.
    \item \textbf{LambdaMART} \cite{burges2010ranknet}: An extension of LambdaRank that employs boosted decision trees and is widely used in production search engines.
\end{itemize}
While effective, these methods require domain-specific training data and do not dynamically adapt to query-specific secondary criteria like recency or diversity.

\subsection{Knowledge Distillation and Hybrid Models}
Dense-sparse hybrid approaches and knowledge distillation techniques combine the strengths of dense embeddings and traditional IR methods:
\begin{itemize}
    \item \textbf{DeepCT} \cite{dai2020deepct}: Enhances sparse term-based retrieval (e.g., BM25) by using dense embeddings to adjust term weights.
    \item \textbf{ANCE} \cite{xiong2020ance}: A dense retrieval model trained with hard negative samples to improve ranking quality.
    \item \textbf{ColBERT} \cite{khattab2020colbert}: Combines token-level interactions with dense embeddings for reranking, offering fine-grained control over relevance scoring.
\end{itemize}
While effective, these methods rely on supervised training and do not offer query-specific adaptability without retraining.

\subsection{Reinforcement Learning for Ranking}
Reinforcement learning (RL) methods optimize ranking policies based on user interactions or feedback:
\begin{itemize}
    \item \textbf{SlateQ} \cite{ie2019slateq}: Optimizes document slates (batches) by balancing multiple objectives, such as relevance and diversity.
\end{itemize}
RL methods are constrained by the need for explicit reward signals, which are not always available in RAG systems.

\subsection{Multi-Objective Optimization (MOO) in IR}
Multi-objective optimization (MOO) frameworks aim to balance competing objectives in ranking:
\begin{itemize}
    \item \textbf{Pareto-Optimal Reranking}: Ensures rankings lie on the tradeoff curve of objectives like relevance, diversity, and recency.
    \item \textbf{Weighted Sum Techniques}: Combines scores for multiple criteria using pre-defined static weights.
\end{itemize}
These methods are limited by the need for static weights, which fail to account for query-specific priorities.

\subsection{Human-Inspired Heuristics}
Heuristic-based approaches explicitly encode human-like priorities:
\begin{itemize}
    \item \textbf{Trust-Aware Ranking}: Prioritizes credible sources, such as peer-reviewed articles or verified authors.
    \item \textbf{Recency-Boosting Search}: Applies temporal decay functions to prioritize recent content unless the query specifies otherwise.
\end{itemize}
While simple to implement, these heuristics are often static and cannot adapt dynamically to individual queries.

\subsection{Active Learning for Multi-Criteria Ranking}
Active learning frameworks iteratively refine rankings based on user feedback:
\begin{itemize}
    \item User-in-the-loop approaches \cite{zheng2010active}: Query users to adjust weights or refine ranking criteria dynamically.
\end{itemize}
These methods introduce latency and are impractical for real-time RAG pipelines.

\subsection{Limitations of Existing Approaches}
The methods discussed above provide valuable insights into multi-criteria and hybrid ranking. However, they often rely on static definitions of ranking criteria, require significant supervision or retraining, and lack the ability to dynamically adapt to query-specific needs. These limitations underscore the need for adaptive, lightweight methods like REBEL, which leverage the flexibility of LLMs without requiring fine-tuning or domain-specific training data.

    \section{Prompts}
    \subsection{Meta Prompt For Reranking Prompt Generator In Two-Turn REBEL}
    \label{sec:meta_prompt}
    \begin{lstlisting}[
    language=Python,
    basicstyle=\ttfamily\small,
    breaklines=true,
    frame=single,
    caption={Default meta prompt used to generate query-dependent reranking prompts. This prompt guides the LLM to create customized reranking instructions that consider both relevance and inferred secondary criteria specific to each query.},
    label={lst:meta_prompt},
    keywordstyle=\color{blue},
    commentstyle=\color{green!60!black},
    stringstyle=\color{purple},
    showstringspaces=false,
    float=false,
    floatplacement=t,
    numbers=left
]
META_PROMPT = '''
You are a prompt generator. You will receive only a user's query as input. Your task is to:

Analyze the user's query and identify additional properties beyond basic relevance that would be desirable for selecting and ranking context documents. These properties should be inferred from the query's subject matter, without the user specifying them. Such properties may include:

Domain appropriateness (e.g., technical accuracy, authoritative sourcing, correctness of information)
Perspective diversity (multiple viewpoints, ideological balance, different theoretical frameworks)
Temporal relevance (up-to-date information, recent data)
Depth of detail and specificity (thorough coverage, multi-faceted analysis, detailed examples)
Trustworthiness, neutrality, impartiality (reliable sources, unbiased accounts)
Reasoning depth or conceptual complexity
Authoritativeness (recognition of reputable experts, institutions, or high-quality sources)
After inferring these properties from the query, produce a final prompt that instructs a large-language model re-ranker on how to:

Take the user's query and a set of candidate documents.
The documents and the query will appear after your instructions in this format: A list of documents is shown below. Each document has a number and a summary. The summaries may indicate the type of source, credibility level, publication date, or the nature of the information. After listing all documents, the user's query will be presented on a single line labeled "Question:". For example: Document 1: <summary of document 1> Document 2: <summary of document 2> ... Document N: <summary of document N> Question: <user's query>
Assign each document a Relevance score (0-10) and scores for each inferred property (0-5).
Compute a weighted composite score for each document. This composite score should not just be used to break ties, but to determine the final ordering. For instance, you may define a formula like: Final Score = Relevance + (Weight1 * Property1) + (Weight2 * Property2) + ... The weights should be specified by you. For example, if you have three properties, you might say: Final Score = Relevance + 0.5*(Property1) + 0.5*(Property2) + 0.5*(Property3) This ensures that documents which strongly exhibit the desired secondary properties can surpass documents with slightly higher relevance but weaker secondary property scores.
Filter out irrelevant documents first. For example, discard any document with Relevance < 3.
Rank all remaining documents by their Final Score (based on the chosen weights).
If two documents end up with the exact same Final Score, you may choose a consistent approach to pick one over the other (e.g., prefer the document with higher authoritativeness).
If no documents meet the relevance threshold, output nothing.
Produce only the final ranked list of chosen documents with their Final Score, in descending order of Final Score. The format for each selected document should be: Doc: [document number], Relevance: [score], where [score] is actually the final score - not the relevance score.
Include no commentary, explanation, or additional text beyond these lines.
Your final prompt should:

Include the user's query verbatim.
Enumerate and define the inferred properties in detail, clearly stating their significance.
Provide the exact scoring rubric for Relevance (0-10) and each inferred property (0-5), explaining what high and low scores mean.
Specify the weighted composite score formula and list the weights for each property.
Give a step-by-step procedure: assign Relevance, assign property scores, discard low-relevance documents, compute Final Scores, sort by Final Score, handle ties if any, then output the final list.
State what to do if no documents qualify (output nothing).
Remind the re-ranker that the documents and query will be shown after this prompt, and that the only acceptable output is the final sorted list of documents and their relevance scores.

At the end of your prompt, you should ALWAYS NO MATTER WHAT include the following:

"Example format: \n"
"Document 1:\n<summary of document 1>\n\n"
"Document 2:\n<summary of document 2>\n\n"
"...\n\n"
"Document 10:\n<summary of document 10>\n\n"
"Question: <question>\n"
"Answer:\n"
"Doc: 9, Relevance: 7\n"
"Doc: 3, Relevance: 4\n"
"Doc: 7, Relevance: 3\n\n"
"Let's try this now: \n\n"
"{context_str}\n"
"Question: {query_str}\n"
"Answer:\n"

Below are 5 k-shot examples demonstrating the required level of detail and explicitness. Each example:

Presents a user query.
Infers multiple properties and explains their relevance.
Provides a scoring rubric for Relevance and the inferred properties.
Defines a weighted composite scoring formula that incorporates Relevance and all secondary properties.
Gives step-by-step instructions for scoring, filtering, ranking, and outputting results.
Explains what to do if no suitable documents remain.
Instructs that the final output should only be lines of the form "Doc: [number], Relevance: [score]" with no extra text.
Example 1 User Query: "How do different countries' tax policies affect income inequality, and what arguments exist from various economic schools of thought?"

Inferred Properties:

Perspective diversity (0-5): Documents that mention or compare multiple economic theories or viewpoints score higher. A high score (5) means it covers several distinct schools of thought. A low score (0) means it is one-dimensional.
Authoritativeness (0-5): Documents from credible economists, reputable research institutes, or peer-reviewed studies score higher. A 5 might be a well-cited academic paper; a 0 might be an anonymous blog post.
Comparative breadth (0-5): Documents discussing tax policies in multiple countries score higher. A 5 means it covers several countries, a 0 means it focuses on just one or does not compare countries at all.
Scoring Rubric: Relevance (0-10): A 10 means the document directly addresses how tax policies influence income inequality and references arguments from different economic viewpoints. A 0 means it is off-topic. Perspective diversity (0-5): Assign based on how many distinct economic perspectives are included. Authoritativeness (0-5): Assign based on credibility and source quality. Comparative breadth (0-5): Assign based on the number of countries or breadth of international comparison.

Weighted Composite Score: Final Score = Relevance + 0.5*(Perspective diversity) + 0.5*(Authoritativeness) + 0.5*(Comparative breadth)

Instructions: After this prompt, you will see: Document 1: <summary> Document 2: <summary> ... Document N: <summary> Question: "How do different countries' tax policies affect income inequality, and what arguments exist from various economic schools of thought?"

Assign Relevance to each document (0-10). Discard documents with Relevance < 3.
For remaining documents, assign Perspective diversity, Authoritativeness, and Comparative breadth (each 0-5).
Compute Final Score as described above.
Sort all remaining documents by Final Score (descending).
If two documents have identical Final Scores, pick consistently, for example by preferring the one with higher Authoritativeness.
If no document remains, output nothing.
Output only: Doc: [number], Relevance: [score] for each selected document, no commentary or explanation, where [score] is actually the final score.

"Example format: \n"
"Document 1:\n<summary of document 1>\n\n"
"Document 2:\n<summary of document 2>\n\n"
"...\n\n"
"Document 10:\n<summary of document 10>\n\n"
"Question: <question>\n"
"Answer:\n"
"Doc: 9, Relevance: 7\n"
"Doc: 3, Relevance: 4\n"
"Doc: 7, Relevance: 3\n\n"
"Let's try this now: \n\n"
"{context_str}\n"
"Question: {query_str}\n"
"Answer:\n"


Example 2 User Query: "What are the latest recommended treatments for chronic lower back pain according to recent medical research?"

Inferred Properties:

Recency (0-5): Higher if the document references recent studies, new clinical guidelines, or up-to-date research (within the last few years). A 5 means it is very recent, a 0 means outdated or no mention of timeliness.
Authoritativeness (0-5): Higher if sourced from reputable medical journals, recognized health organizations, or consensus guidelines.
Specificity (0-5): Higher if it focuses specifically on chronic lower back pain treatments. A 5 means it precisely addresses chronic lower back pain, a 0 means it only vaguely mentions pain or general treatments without specificity.
Scoring Rubric: Relevance (0-10): A 10 means the document explicitly discusses current recommended treatments for chronic lower back pain based on recent research. A 0 means off-topic. Recency (0-5) Authoritativeness (0-5) Specificity (0-5)

Weighted Composite Score: Final Score = Relevance + 0.5*(Recency) + 0.5*(Authoritativeness) + 0.5*(Specificity)

Instructions: After this prompt: Document 1: <summary> ... Document N: <summary> Question: "What are the latest recommended treatments for chronic lower back pain according to recent medical research?"

Assign Relevance. Exclude Relevance < 3.
Assign Recency, Authoritativeness, Specificity.
Compute Final Score.
Sort by Final Score.
If tied, choose consistently (e.g., prefer higher Authoritativeness).
If none remain, output nothing.
Output only lines like: Doc: X, Relevance: Y, where Y is actually the final score.

"Example format: \n"
"Document 1:\n<summary of document 1>\n\n"
"Document 2:\n<summary of document 2>\n\n"
"...\n\n"
"Document 10:\n<summary of document 10>\n\n"
"Question: <question>\n"
"Answer:\n"
"Doc: 9, Relevance: 7\n"
"Doc: 3, Relevance: 4\n"
"Doc: 7, Relevance: 3\n\n"
"Let's try this now: \n\n"
"{context_str}\n"
"Question: {query_str}\n"
"Answer:\n"


Example 3 User Query: "How did the policies of Emperor Qin Shi Huang shape the political and cultural landscape of ancient China?"

Inferred Properties:

Historical depth (0-5): Higher if it provides detailed historical context, dates, and direct evidence. A 5 is richly detailed, a 0 is very superficial.
Perspective range (0-5): Higher if it references multiple historians or scholarly opinions. A 5 means multiple perspectives, a 0 is one-sided.
Cultural/political detail (0-5): Higher if it addresses both political structures and cultural changes. A 5 is comprehensive, a 0 is minimal detail.
Scoring Rubric: Relevance (0-10): A 10 means it explicitly discusses Qin Shi Huang's policies and their impact on both political and cultural aspects of ancient China. Historical depth (0-5) Perspective range (0-5) Cultural/political detail (0-5)

Weighted Composite Score: Final Score = Relevance + 0.5*(Historical depth) + 0.5*(Perspective range) + 0.5*(Cultural/political detail)

Instructions: After this prompt: Document 1: <summary> ... Document N: <summary> Question: "How did the policies of Emperor Qin Shi Huang shape the political and cultural landscape of ancient China?"

Assign Relevance, discard < 3.
Assign Historical depth, Perspective range, Cultural/political detail.
Compute Final Score.
Sort by Final Score.
Tie-break by preferring more historically authoritative perspectives if still tied.
If none qualify, output nothing.
Only output: Doc: [number], Relevance: [score], where [score] is actually the final score.

"Example format: \n"
"Document 1:\n<summary of document 1>\n\n"
"Document 2:\n<summary of document 2>\n\n"
"...\n\n"
"Document 10:\n<summary of document 10>\n\n"
"Question: <question>\n"
"Answer:\n"
"Doc: 9, Relevance: 7\n"
"Doc: 3, Relevance: 4\n"
"Doc: 7, Relevance: 3\n\n"
"Let's try this now: \n\n"
"{context_str}\n"
"Question: {query_str}\n"
"Answer:\n"


Example 4 User Query: "What are the main differences between various machine learning frameworks like TensorFlow, PyTorch, and Scikit-learn?"

Inferred Properties:

Technical accuracy (0-5): Higher if the document correctly and specifically describes features, performance characteristics, or typical uses. A 5 means very accurate and specific.
Comparative breadth (0-5): Higher if the document compares multiple frameworks directly, ideally all three. A 5 means it covers all three well, a 0 means it only mentions one.
Authoritativeness (0-5): Higher if citing official documentation, known ML experts, or reputable evaluation sources.
Scoring Rubric: Relevance (0-10): A 10 means the document explicitly compares these ML frameworks in detail. Technical accuracy (0-5) Comparative breadth (0-5) Authoritativeness (0-5)

Weighted Composite Score: Final Score = Relevance + 0.5*(Technical accuracy) + 0.5*(Comparative breadth) + 0.5*(Authoritativeness)

Instructions: After prompt: Document 1: <summary> ... Document N: <summary> Question: "What are the main differences between various machine learning frameworks like TensorFlow, PyTorch, and Scikit-learn?"

Assign Relevance, exclude < 3.
Assign Technical accuracy, Comparative breadth, Authoritativeness.
Compute Final Score.
Sort by Final Score.
Tie-break by preferring documents that are more authoritative or have greater comparative breadth.
If none remain, output nothing.
Output only lines like: Doc: [number], Relevance: [score], where [score] is actually the final score.

"Example format: \n"
"Document 1:\n<summary of document 1>\n\n"
"Document 2:\n<summary of document 2>\n\n"
"...\n\n"
"Document 10:\n<summary of document 10>\n\n"
"Question: <question>\n"
"Answer:\n"
"Doc: 9, Relevance: 7\n"
"Doc: 3, Relevance: 4\n"
"Doc: 7, Relevance: 3\n\n"
"Let's try this now: \n\n"
"{context_str}\n"
"Question: {query_str}\n"
"Answer:\n"

Example 5 User Query: "What are the arguments for and against universal basic income in modern economies?"

Inferred Properties:

Balance of perspectives (0-5): Higher if the document presents both pro and con arguments. A 5 means thorough coverage of both sides.
Reasoning depth (0-5): Higher if it explains the rationale behind arguments, providing logic or evidence.
Authoritativeness (0-5): Higher if referencing economists, studies, or policy analyses from reputable sources.
Scoring Rubric: Relevance (0-10): A 10 means it clearly discusses UBI arguments both for and against. Balance of perspectives (0-5) Reasoning depth (0-5) Authoritativeness (0-5)


Weighted Composite Score: Final Score = Relevance + 0.5*(Balance of perspectives) + 0.5*(Reasoning depth) + 0.5*(Authoritativeness)

Instructions: After prompt: Document 1: <summary> ... Document N: <summary> Question: "What are the arguments for and against universal basic income in modern economies?"

Assign Relevance, discard < 3.
Assign Balance of perspectives, Reasoning depth, Authoritativeness.
Compute Final Score.
Sort by Final Score.
If tied, prefer documents with higher reasoning depth or authoritativeness.
If none remain, output nothing.
Output only: Doc: [number], Relevance: [score], where [score] is actually the final score.

"Example format: \n"
"Document 1:\n<summary of document 1>\n\n"
"Document 2:\n<summary of document 2>\n\n"
"...\n\n"
"Document 10:\n<summary of document 10>\n\n"
"Question: <question>\n"
"Answer:\n"
"Doc: 9, Relevance: 7\n"
"Doc: 3, Relevance: 4\n"
"Doc: 7, Relevance: 3\n\n"
"Let's try this now: \n\n"
"{context_str}\n"
"Question: {query_str}\n"
"Answer:\n"


Follow these examples as a template for your final prompt. For any new user query, do the following:

Include the user's query verbatim.
Infer the relevant secondary properties and define them clearly.
Give a scoring rubric for Relevance and each property.
Specify a weighted composite score formula that combines Relevance and the properties.
Provide step-by-step instructions: assign scores, filter out irrelevant documents, compute Final Score, sort by Final Score, handle ties, and if none qualify, output nothing.
Instruct the re-ranker to output only the final list of documents and their Relevance scores, with no extra commentary.
Now, here is the user's query:

[USER QUERY]
'''
\end{lstlisting}
    \subsection{One-Turn Multi-Criteria Reranking Prompt}
    \label{sec:static_reranking_prompt}
    \begin{lstlisting}[
    language=Python,
    basicstyle=\ttfamily\small,
    breaklines=true,
    frame=single,
    caption={Default prompt used to rerank documents with a diverse set of multiple criteria.},
    label={lst:reranking_prompt},
    keywordstyle=\color{blue},
    commentstyle=\color{green!60!black},
    stringstyle=\color{purple},
    showstringspaces=false,
    float=false,
    floatplacement=t,
    numbers=left
]
DEFAULT_CHOICE_SELECT_PROMPT_TMPL = '''
            You are a re-ranking system. Your task is to analyze a user's query and a set of candidate documents, assign scores based on specified properties, and output the final ranking of documents.

            **Inferred Properties**

            1. **Depth of Content (0-5):**
            - Higher scores indicate thorough detail and comprehensive coverage of the topic.
            - A "5" is exceptionally in-depth with multiple facets addressed; a "0" is very superficial.

            2. **Diversity of Perspectives (0-5):**
            - Higher scores indicate that multiple viewpoints or angles are represented.
            - A "5" means it engages with a variety of perspectives or sources; a "0" means it is entirely one-sided.

            3. **Clarity and Specificity (0-5):**
            - Higher scores indicate that the document presents information clearly and addresses the query with precise, unambiguous detail.
            - A "5" means it is highly specific and clear, while a "0" means it is vague or overly general.

            4. **Authoritativeness (0-5):**
            - Higher scores indicate reputable sources, expert authorship, or recognized credibility.
            - A "5" might be an extensively cited academic work or an official standard; a "0" would be an unknown or dubious source.

            5. **Recency (0-5):**
            - Higher scores indicate that the document references recent studies, data, or developments.
            - A "5" means it is very current and up-to-date; a "0" means it is outdated or does not reference any time-sensitive information.

            **Scoring Rubric**

            - **Relevance (0-10):**
            - A "10" means the document directly addresses the user's query, covering the key subject comprehensively.
            - A "0" means it is completely off-topic.

            - **Depth of Content (0-5):** Based on how detailed or thorough the document is.
            - **Diversity of Perspectives (0-5):** Based on how many viewpoints or angles are presented.
            - **Clarity and Specificity (0-5):** Based on how clear and precise the document is.
            - **Authoritativeness (0-5):** Based on source credibility or recognized expertise.
            - **Recency (0-5):** Based on how up-to-date the document is.

            **Weighted Composite Score**
            Final Score = Relevance + 0.5*(Depth of Content) + 0.5*(Diversity of Perspectives) + 0.5*(Clarity and Specificity) + 0.5*(Authoritativeness) + 0.5*(Recency)

            **Instructions**
            1. Assign Relevance to each document on a scale of 0-10. Discard any document with Relevance < 3.
            2. For the remaining documents, assign scores for:
            - Depth of Content (0-5)
            - Diversity of Perspectives (0-5)
            - Clarity and Specificity (0-5)
            - Authoritativeness (0-5)
            - Recency (0-5)
            3. Compute each document's Final Score using the formula above.
            4. Sort the documents by their Final Score in descending order.
            5. If two documents end up with the same Final Score, prefer the one with higher Authoritativeness (or apply another consistent tie-breaking rule).
            6. If no documents remain after filtering for Relevance, output nothing.
            7. Output only the list of selected documents with their Relevance scores, in this format (no extra text or commentary), where [score] is actually the Final Score and NOT the relevance score.:
            ```
            Doc: [document number], Relevance: [score]
            ```

            **Example format:**
            ```
            Document 1:
            <summary of document 1>

            Document 2:
            <summary of document 2>

            ...

            Document 10:
            <summary of document 10>

            Question: <question>
            Answer:
            Doc: 9, Relevance: 7
            Doc: 3, Relevance: 4
            Doc: 7, Relevance: 3

            Let's try this now:

            {context_str}
            Question: {query_str}
            Answer:
            ```
            '''
\end{lstlisting}

\section{Rationale For Choosing Answer Similarity Over Alternative Answer Quality Metrics}
\label{sec:alternative_answer_quality_evaluation_metrics}
Alternative answer quality evaluation approaches often fall short in multiple ways. Surface-level metrics like ROUGE \citep{lin2004rouge} and BLEU \citep{papineni2002bleu} can be misled by superficial text matching, marking responses as different even when they convey identical meanings through different words. While more sophisticated metrics like BERTScore \citep{zhang2020bertscore} move beyond n-gram matching by using contextual embeddings, the semantic relationships captured by these embedding distances remain opaque, and their uncalibrated scores lack clear interpretability - a 0.8 BERTScore doesn't map to any intuitive measure of answer quality. Many evaluation frameworks compound these issues by relying on complex scoring mechanisms or multiple separate metrics that require careful tuning of thresholds and weights \citep{karpukhin2020dense, khattab2020colbert, lewis2020retrieve}, making system comparisons difficult and obscuring what matters most - whether the system produces answers that convey the intended meaning. Our answer similarity metric addresses these limitations through a deliberately streamlined approach, directly asking an LLM to rate semantic similarity between generated and reference answers on a 0-5 scale. This straightforward assessment focuses on the core question: does the generated answer convey the same meaning as the reference? By reducing evaluation to this fundamental comparison and moving beyond both syntactic similarities and abstract embedding spaces, we enable a clearer assessment of whether a RAG system is achieving its fundamental goal: producing answers that convey the same meaning as high-quality reference responses, regardless of exact wording.

\section{Future Work}
\label{sec:mt_improvements}

Given that both versions of REBEL use Chain-of-Thought prompting, and our two-turn version uses multi-turn techniques, several promising avenues for improvement emerge from recent advances in Chain-of-Thought and multi-turn prompting. 

\subsection{Enhancing Chain-of-Thought Prompting in Both Variants}
\subsubsection{Static Multi-Criteria Reranking Prompt Improvements}
Several strategies could enhance our fixed reranking prompt:

\begin{itemize}
    \item \textbf{Empirical Weight Tuning:} Following \citet{pryzant2023automatic}, we could systematically evaluate different weightings for our five fixed criteria (depth, diversity, clarity, authoritativeness, and recency) across diverse query types. This could help identify optimal default weights that generalize well across different scenarios.
    
    \item \textbf{Criteria Definition Refinement:} Drawing from \citet{zhou2022least}, we could break down each criterion into more precise sub-components with clearer scoring guidelines. For instance, "depth" could be decomposed into measurable aspects like "number of distinct concepts covered" and "level of technical detail."
    
    \item \textbf{Scoring Rubric Optimization:} Inspired by \citet{wang2022self}, we could generate multiple candidate rubrics for each criterion, evaluate their effectiveness through controlled experiments, and synthesize the most reliable scoring guidelines. This could improve the consistency and interpretability of our scoring system.
    
    \item \textbf{Cross-Criteria Interaction Analysis:} Using techniques from \citet{yao2023tree}, we could explore how different criteria interact and potentially modify our scoring formula to account for these interactions. For example, we might discover that high diversity scores are more valuable when combined with high authoritativeness.
\end{itemize}

\subsubsection{Meta Prompt Improvements}
For our meta prompt that generates query-dependent reranking instructions, we identify several potential enhancements:

\begin{itemize}
    \item \textbf{Example Diversification:} Following \citet{wei2022chain}, we could expand our k-shot examples to cover a broader range of query types and domains, helping the prompt generator better adapt to diverse information needs. \citet{min2022rethinking} suggest this could be further enhanced by selecting examples that specifically target edge cases and challenging scenarios.
    
    \item \textbf{Dynamic Weight Assignment:} Inspired by \citet{fu2023complexity}, we could enhance the meta prompt's ability to assign appropriate weights to inferred criteria based on query complexity characteristics. This might involve providing explicit guidelines for weight selection based on query features like complexity, domain, or intended use as demonstrated in \citet{zhang2023automatic}.
    
    \item \textbf{Output Format Optimization:} Following \citet{mishra2023decomposed}, we could refine how the generated reranking prompts structure their scoring guidelines and instructions, potentially incorporating more explicit step-by-step breakdowns to improve clarity and consistency.
\end{itemize}

These improvements could enhance both the reliability of our fixed criteria evaluation and the adaptability of our query-dependent approach. Future work should systematically evaluate these modifications to identify which combinations yield the most robust and effective reranking strategies.

\subsection{Multi-Turn Dialogue Advancements}
Several promising avenues for improving our two-turn emerge from recent advances in multi-turn dialogue systems and iterative prompting. These are outlined as follows:
\begin{itemize}
    \item \textbf{Progressive Refinement:} Following \citet{diao2023progressive}, we could implement a step-wise refinement process where each turn builds upon and refines the criteria identified in previous turns. This could help ensure more robust and comprehensive criteria identification.
    
    \item \textbf{Recursive Prompting:} Inspired by \citet{zhou2023recursive}, we could expand our two-turn approach into a recursive structure where each level of criteria inference informs and refines the next. This would enable the system to explore and evaluate criteria at multiple levels of granularity.
    
    \item \textbf{Structured Turn Taking:} Adapting the approach of \citet{wu2023structured}, we could organize the multi-turn dialogue into distinct phases for criteria identification, evaluation, and refinement. This structured approach could be particularly valuable for complex queries requiring multiple types of criteria.
    
    \item \textbf{Adaptive Turn Iteration:} Drawing from \citet{jung2023adaptive}, we could dynamically adjust the number and nature of turns based on query complexity, allowing for more efficient and targeted criteria inference.
\end{itemize}
\end{document}